\documentstyle[aps,preprint,prb,epsfig]{revtex}

\draft

\tighten

\title{Memorization of short-range potential fluctuations in Landau levels}

\author{Andrei Manolescu$^1$ and Vidar Gudmundsson$^2$}

\address{
$^1$Institutul Na\c{t}ional de Fizica Materialelor, C.P. MG-7 
Bucure\c{s}ti-M\u{a}gurele, Rom\^ania,\\
$^2$Science Institute, University of Iceland, Dunhaga 3, IS-107 Reykjavik, 
Iceland}

\begin{document}

\maketitle

\begin{abstract} 

We calculate energy spectra of a two-dimensional electron system in a
perpendicular magnetic field and periodic potentials of short periods.
The Coulomb interaction is included within a screened Hartree-Fock
approximation.  The electrostatic screening is poor and the exchange
interaction amplifies the energy dispersion. We obtain, by numerical
iterations, self-consistent solutions that have a hysteresis-like
property.  With increasing amplitude of the external potential the
energy dispersion and the electron density become periodic, and they
remain stable when the external potential is reduced to zero. We explain
this property in physical terms and speculate that a real system
could memorize short-range potential fluctuations after the potential
has been turned off.

\end{abstract}

\pacs{71.45.Gm,71.70.Di,73.20.Dx}

\narrowtext

It is well known that in a two-dimensional electron system (2DES), in a
strong perpendicular magnetic field and for low temperatures, the
exchange component of the Coulomb interaction enhances the spin
splitting of the Landau levels when the Fermi level is in a spin gap,
i.\ e.\ for odd filling factors.\cite{Ando74:1044}  The gap enhancement
also occurs for even filling factors, and the energy interval between two
adjacent (even spin degenerate) Landau levels can exceed the cyclotron
energy.  More generally, in the presence of an external potential the
energy dispersion imposed on the Landau levels may be amplified due to
the exchange interaction.\cite{Manolescu95:1703}   The common mechanism
of the exchange enhancement, either of the energy gap or of the energy
dispersion, is the energy decrease of any occupied state by a {\em
negative} amount corresponding to the exchange component.  The exchange
interaction behaves like an attractive short-range potential within
distances of the order of the magnetic length $l=\sqrt{\hbar/(eB)}$.
The exchange energy of a one-particle state may be of the order  $-e^2/(\kappa
l)$, where $\kappa$ is the dielectric constant of the semiconductor host
of the 2DES.  For GaAs systems ($\kappa=12.4$, $m_{\mbox{eff}}=0.067$,
$g_{\mbox{eff}}=-0.44$) this energy is comparable to the cyclotron energy,
$\hbar eB/m_{\mbox{eff}}$, for $B$ about 5 -- 10 T, i.\  e.\ in the range
5 -- 15 meV.

The essential difference between the homogeneous and the inhomogeneous
2DES is the action of the electrostatic screening.  In the homogeneous
system the screening of the electron-electron interaction, by the 2DES
itself, reduces the exchange effects for non-integer filling factors,
when the density of states at the Fermi level (DOSF) is finite.  In the
inhomogeneous system the DOSF decreases due to the energy dispersion,
the screening abilities diminish, and the exchange becomes stronger.
However, the classical, repulsive, Hartree part of the Coulomb
interaction, which is absent in the homogeneous case, tends to suppress the
long-range fluctuations of the charge density and to flatten the energy
dispersion.  Therefore in the presence of an external electrostatic
potential, the energy dispersion of the Landau levels is a complicated
result of combined and opposite exchange and direct contributions
controlled by the potential steepness at the magnetic length scale.
Calculations for periodic potentials of periods much longer than $l$,
within the screened Hartree-Fock approximation (SHFA), have shown Landau
bands typically much narrower than the external potential amplitude, i.\
e.\  strong screening effects, but still a considerable exchange
influence.\cite{Manolescu97:9707}   In this paper we will be interested
in the opposite limit, of weak (in amplitude), but steep electrostatic
potentials, for which one may expect a much weaker screening, but a much
stronger exchange.

Recent experiments show that the spin splitting of the Shubnikov-de
Haas peaks in short-period (30 nm) modulated 2DES has
an abrupt onset when the magnetic field increases.\cite{Petit97:225}
This agrees with the Hartree-Fock calculations for such a system.
Increasing the magnetic field the length $l$ decreases, and so does the
potential gradient experienced by the electrons in low-energy states.
Therefore the wide Landau bands, with strong exchange broadening, and thus
with hidden spin splitting, change suddenly into narrow bands,
but now with a strong enhancement of the spin gap.\cite{Manolescu95:1703}
According to this interpretation the experiment confirms that exchange
effects strongly influence the energy dispersion.  The abrupt transition
from spin unpolarized to spin polarized states has been initially
predicted for narrow quantum wires.\cite{Kinaret90:11768}
A similar transition has been discussed for the edge states, as a direct
implication of the single-particle-energy dispersion: for soft edges the
edge channels are spatially spin split, but they rapidly collapse and
become unpolarized with increasing edge steepness.\cite{Dempsey93:3639}

We take all these results as indications that the exchange interaction
plays an important role in the 2DES in the presence of a strong
potential gradient.  Our main result is that in a weak periodic
electrostatic potential of a short period, for a strong magnetic field,
the width of the Landau bands may be {\em larger} than the potential
amplitude, and {\em robust} to amplitude fluctuations.  The Landau-band
structure, together with the density modulation induced in the 2DES,
with the period of the external potential, can survive after the
potential has been turned off.  We diagonalize the Hamiltonian of the
2DES in the periodic potential $V_x\cos(K_x x) + V_y\cos(K_y y)$,
within the SHFA, following a numerical iterative process.  We consider
both 1D and 2D potentials, where in the former case $V_y=0$ and $K_y=0$.
We shall discuss first the results for a 1D potential.  

The Coulomb
potential in the exchange term of the effective Hamiltonian is 
$u({\bf q})=2\pi e^2/[\kappa q\epsilon (q)]$, $\epsilon (q)$ 
being the static dielectric function of the 2DES, calculated
self-consistently with the DOSF, and also with the (less important in
our regime) wave-functions polarization.\cite{Ando74:1044}   We assume
a finite temperature $T$, and also a weak disorder broadening of the
energy spectrum.  The latter we take into account by the familiar
ansatz that the spectral function is a Gaussian of width $\Gamma$.
All the details of the calculation have been presented in a recent
publication.\cite{Manolescu97:9707}   We consider a modulation period
$a_x\equiv 2\pi/K_x=50$ nm, and a modulation amplitude arbitrarily
fixed in the interval 1.2 meV $<V_x<8.5$ meV in the first part of
the calculation, until a stable solution is obtained, and then we put
$V_x=0$, and continue the iterations until the solution is again stable.
The final state remains periodically modulated and satisfies the
equations of the SHFA, {\em in the absence} of any external potential.
In Fig.\ 1 we show the final state: the first four spin split Landau 
bands in the first Brillouin zone, $0\leq\xi < a_x$, where $\xi$ is 
the center coordinate, and one unit cell of the remanent density 
modulation.  The initial potential drives the system into the
modulated state and determines the spatial period, which is thus
\lq\lq memorized\rq\rq\ by the 2DES after the potential is turned off.
The final solution is independent of the initial potential amplitude,
within the specified interval.

According to the previous discussion, we identify a quantum screening
mechanism, acting on the exchange interaction, and the classical Hartree
screening, directly coupling the induced potential to the electron
density.  Both mechanisms are inhibited in the state displayed in Fig.\ 1.
The steep and stable energy dispersion is the self-consistent result of
a strong exchange enhancement and a low DOSF.  The Hartree mechanism is
expected to restore the homogeneous state after turning the external
potential off.   But, due to the short period, even if the induced
potential is considerable, the classical effect is weak and far from the
linear regime.  We believe the inhibition of the electrostatic forces
is crucial.  Neglecting the Hartree term remanent modulations can be
obtained for any period.  Within our SHFA we could obtain them only up
to $a_x=80$ nm, while for longer periods the system relaxes into the
homogeneous state.

The unusual property of the state of Fig.\ 1 is the negative dielectric
response. The amplitude of the external potential can be exceeded in
absolute value by the amplitude of the induced potential.  Therefore the
total potential (external plus induced) can have maxima where the electron
density has maxima.  Such over-screening states are known for other
(classical or quantum) systems,\cite{Kirzhnits} the stability being
always ensured by additional forces.  In our case this role is played
by the short-range attractive exchange interaction.

In the standard HFA, i.\ e.\  without screening the exchange term, for
$K_xl\ll 1$ spontaneous charge-density wave (CDW) instabilities may occur,
known for a long time, \cite{Fukuyama79:5211}  with an {\em intrinsic}
period determined only by $l$ and material constants, and which are
considered artifacts of the HFA.  We want to stress that this type of
instability is absent in our SHFA.  A new type of CDW ground-states have
been recently obtained in a SHFA when many Landau levels are occupied.
\cite{Fogler}  In that regime the screening is stronger than in ours,
dominated by the wave-functions polarization cumulated from all the
occupied levels, and therefore our memorized modulation is no longer
stable.  However our effect is by no means restricted to the Landau level
with $n=0$ or to very strong magnetic fields, see below.  We could also
obtain it in the Landau levels with $n=1$ and $n=2$.

In Fig.\ 2 we show, very schematically, what we consider to be the
essence of the problem.  Suppose we act on the 2DES with a step
potential, $\phi_0\theta(-x)$.  Assume also all the electrons
are in the lowest Landau level. Ignoring what happens at the
boundaries of the system, we can imagine the Landau level divided
into a fully occupied and an empty region, and the energy dispersion
concentrated in the interval $\mid\xi-\xi_F\mid<l/2$.  The exchange
energy of a single particle state is given by the expression
\cite{Ando74:1044,MacDonald86:2681,Manolescu95:1703} 
$E_{\xi}=-e^2 \nu_{\xi}\sqrt{\pi/2}/(\kappa l)$, where $\nu_{\xi}$ 
reduces to the filling factor of the one-particle states if the center
coordinate is in a region with no energy dispersion, i.\ e.\  0 or 1.
In the Landau gauge the wave functions  outside the transition region are
Gaussians, ${\exp}[-(x-\xi)^2/2l^2]$.  In the transition region, due to
the translational invariance along the $y$ direction, the perturbation
does not mix the center coordinates, but only the Landau levels,
such that the wave functions still have a localized character.  For high
magnetic fields the Landau level mixing weakens (high energy separation)
and $l$ decreases.  The width of the transition region reduces to $l$,
and $\nu_{\xi}$ becomes the Fermi function.  The electron density
must follow a similar step-like profile, and in combination with a
positive uniform background, positive and negative charges separate.
The induced electric field is thus concentrated in the transition region
only, and vanishes outside.  

The essential aspect in this idealized situation is that the exchange
energy does not depend explicitly on $\phi_0$.  The exchange action is
reduced to pushing the occupied energy levels even more below the Fermi
energy, as soon as a sufficiently strong energy gradient has been imposed.
The energy gradient is thus amplified to the value indicated in Fig.\
2, since due to the low DOSF the electrons are not able to screen the
exchange interaction.  Therefore, by suppressing the external potential
the state remains stable.  For $B\to\infty$ the induced electric field
acts on a vanishing number of particles, and is not capable to restore
the homogeneous state.  However, for a very weak $\phi_0$ the imposed
energy dispersion may be insufficient for being further subjected to the
exchange enhancement. In this situation the 2DES would still behave like
a homogeneous system, with complex, nearly perfect screening properties,
involving also correlation energy.  We thus have to imagine the transition
occurring only when $e\phi_0$ exceeds a certain threshold energy $V_t$.
Above this threshold the step potential produces the {\em shearing effect}
depicted in Fig.\ 2.

For the 1D periodic potential used for Fig.\ 1, and for the temperature
$T=0.9$ K, we obtain $V_t=1.2$ meV.  With $V_x<V_t$ in the initial
stage of the calculation the solution converges to the homogeneous
state after putting $V_x=0$.  In terms of the energy per particle,
$E/N$, this sort of hysteresis effect is described in Fig.\ 3.  The solid
line shows the evolution of the energy with increasing external
modulation amplitude (when it is maintained through all the
iterations).  For $V_x<V_t$ the system is
in the linear screening regime, where the Landau bands are very narrow
and the density profile is almost flat, with nearly no difference with
respect to the Hartree approximation.  For $V_x=V_t$ the screening is
suddenly suppressed and the exchange interaction lowers the energy
abruptly. The Landau bands look qualitatively like in Fig.\ 1, up to the
second discontinuity, at $V_x=8.5$ meV.  Here the spin split bands rapidly
overlap at the Fermi level and thus the exchange energy is again screened.
This situation is analog to that detected in the experiments by Petit et
al., \cite{Petit97:225} now with a fixed magnetic field.  Suppressing here
the external potential our solution relaxes towards the homogeneous state.
However, by reducing it gradually, and thus making weaker numerical
perturbations, the system evolves back into the states with low DOSF,
and then continues along the dotted line down to $V_x=0$.

$V_t$ increases fast with the temperature, as shown by the dash-dotted
line of Fig.\ 3.  The reason is that the slope of the Landau band at
the Fermi level decreases with increasing $T$,  and thus again DOSF
increases and the strong exchange effect vanishes.  However, when that
slope exceeds a critical value the temperature has practically no more
influence.\cite{Dempsey93:3639} These are the situations with low
DOSF, i.\ e.\ the interval between the energy jumps.  Nevertheless,
the modulated solutions are now less robust
if we reduce $V_x$ during the calculation, such that we could obtain
hysteresis effect only by slowly (adiabatically) reducing $V_x$
below $V_t$, and we could not reach $V_x=0$ in all cases.

The hysteresis effect totally disappears for temperatures slightly above
1.3 K.  The effect of the disorder parameter is similar to that of
the temperature, since the occupation numbers of the single-particle
states depend on $\Gamma$.  In all the calculations we have fixed
$\Gamma=0.13$ meV.  One could also define a critical disorder parameter,
but we only restrict ourselves to stating that slight inhomogeneities
do not destroy the hysteresis solution.  Another situation with strong
screening and no memory effect is when the filling factor is close to
an integer value.  The DOSF may be above the critical value due to the
proximity of a van Hove singularity.  The screening effects also
increase when the modulation period increases, and again $V_t$ grows
until the low-DOSF states are no longer stable.  Obviously, with
decreasing period below the magnetic length all the potential 
oscillations are averaged and the exchange interaction becomes again 
unimportant.

In the numerical calculation we have made two simplifications in the
screening of the exchange interaction.  First, we have used a static
screening approximation. The dynamic effects can be estimated within
the Coulomb hole approximation which successfully explained experimental
luminescence spectra.\cite{Katayama89:97}  But the dynamic effects {\em
increase} the exchange interaction, except only for integer filling
factors.  Second, we have considered a dielectric function calculated
as in the case of a homogeneous system, 
by taking an average filling factor into
account.\cite{Manolescu97:9707}  This is in principle correct for a weakly
modulated system, while here one should consider a dielectric matrix.
Of course, in this respect our treatment is not fully self-consistent.
Such a calculation would be extremely difficult, but we argue that the
modulated system treated that way would posses a certain rigidity which
makes it less sensitive to electric perturbations, and reduces even the
screening effect on the exchange interaction.  We thus believe that,
in fact, the exchange effects are even {\em underestimated} by our
approximations.

Another aspect is the energy comparison between the periodic states with
no external potential and the homogeneous state.  In the SHFA the periodic
states have lower energies, Fig.\ 3.  But of course this comparison
is irrelevant.  For inhomogeneous states SHFA or HFA could be realistic
approximations, but not for homogeneous states, where correlation energy
is known to be important.  Attempting to include correlations beyond
the present SHFA only in the homogeneous states, the comparison would be
further vitiated by the use of two different Hamiltonians for the same
system, with and without the external potential.  Nevertheless, we have
compared with the energy of the homogeneous state for the lowest, spin-less
Landau level, estimated by Fano and Ortolani,\cite{Fano88:8179} and we
have found our periodic states having higher energies.  For instance
for $B=13$ T our modulated solution in the lowest (spin up) Landau
level gives $E/N=3.88$ meV whereas their formula gives $E/N=2.35$ meV.
This result suggests that the states with remanent density modulations
-- if they exist -- could be meta-stable.  Accurate calculations
of $E/N$ are not available for higher Landau levels, so our comparison
is extremely limited.  Since in our calculation we do not use any free
internal parameter of the system we are also not able to investigate
energy minima.  All we can do is to check the stability numerically with
respect to charge fluctuations, screening, or thermal effects.

We want to mention that the spin plays here no fundamental role.  The spin
splitting only allows us to obtain a sufficiently low DOSF and thus a
sufficiently strong exchange effect.  We can obtain remanent modulations
with two Landau bands at the Fermi level by reducing $\kappa$ and hence
by increasing the exchange energy, see Fig. 2.  We can also obtain
our exchange effects in the unrestricted Hartree-Fock approximation,
where the electron spin is not restricted to have only components
parallel or anti parallel to the external 
magnetic field.\cite{Palacios5760:1994}

For a 2D external potential, due to the technical difficulties imposed
by the magnetic flux commensurability condition, we do not screen the
exchange interaction in a self-consistent manner.  Instead, we consider a
constant Thomas-Fermi wave vector $q_{TF}=0.2/l$ in the Coulomb potential.
The only qualitative difference with respect to the self-consistent
calculations is that the threshold potential $V_t$ is drastically
reduced.  For a symmetric initial potential, $V_x=V_y$ and periods
$a_x=a_y$, the final state preserves the same 2D periodicity, which
is just the isotropic version of the results obtained in the 1D case.
For an asymmetric initial potential, the final state may have only a
1D symmetry.  Such an example is shown in Fig.\ 5, where $a_x=50$ nm and
$a_y=70$ nm.  The energy dispersion is described by the quantum numbers
$\theta_{1,2}$ given by the momenta in the first magnetic Brillouin zone.
\cite{Gudmundsson95:16744}  Here we have chosen a magnetic flux equal
to one flux quanta per unit cell, such that the sub-band (Hofstadter)
splitting is absent.  Since the intersection of the Fermi surface with
the Landau band is an open contour, only the dispersion in one direction
captures single-particle states below the Fermi level, and only the period
for this direction is memorized, which here is $a_y$.   Therefore  new
types of threshold-amplitude conditions are present with an initial 2D
potential, in order to keep in the final electron density the periodicity
in the other or in both direction.  The initial amplitudes $V_{x,y}$
for Fig.\ 5 have been independently varied within the interval 1 -- 2 meV.

In conclusion, we have found a hysteresis property of our periodic
numerical solution with respect to the external potential amplitude.
We have explained it in physical terms, the key elements being the strong
exchange interaction and the poor screening in the presence of steep
potentials.  Therefore we speculate on our numerical results predicting
that acting on a real system with an electrostatic potential with strong
gradient at the magnetic length scale it is possible to get shearing
of the electron density which remains stable after reducing or even
switching off the potential, thus memorizing its short-range fluctuations.
Periodic potentials with controlled amplitudes can be obtained with metallic
strip gate.  The electrical resistance measured perpendicular
to the strips should increase in the presence of a gate voltage,
and this could be one possibility of probing the hysteresis effect.
Another possibility could be the potential investigation by one of the
electrostatic imaging methods just under development.

This research was supported by the Icelandic Natural Science Foundation,
the University of Iceland Research Fund, and by the International Centre
for Theoretical Physics, Trieste, Italy, within the Associateship Scheme.
Stimulating discussions with Rolf Gerhardts, Daniela Pfannkuche, and
Behnam Farid are acknowledged.

\newpage

\begin{figure} 
\vspace{1cm}
\epsfxsize 12cm 
\begin{center}
\epsffile{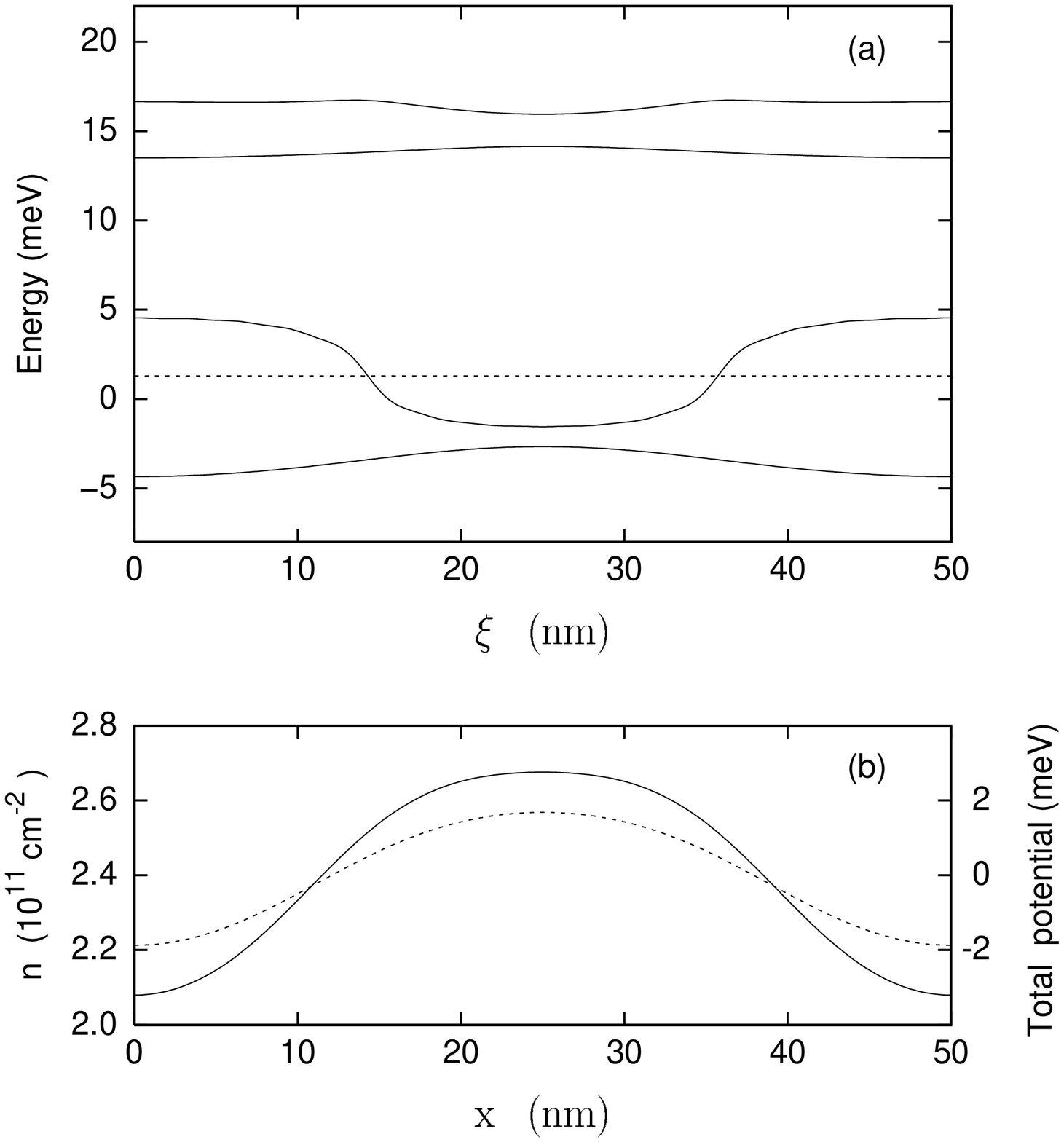}
\vspace{-1cm}
\end{center}
\caption{Final state after the supression of the external potential. 
The parameters are $B=7$ T, $T=0.9$ K, $\Gamma=0.13$ meV,
and the average electron density $n_s=2.4\times 10^{11}$ cm$^{-2}$.
(a) Energy spectrum and the Fermi level (dotted line). 
(b) Electron density, full line, and the internal electrostatic
potential, dotted line.}
\end{figure}
\begin{figure}
\epsfxsize 10cm
\begin{center}
\epsffile{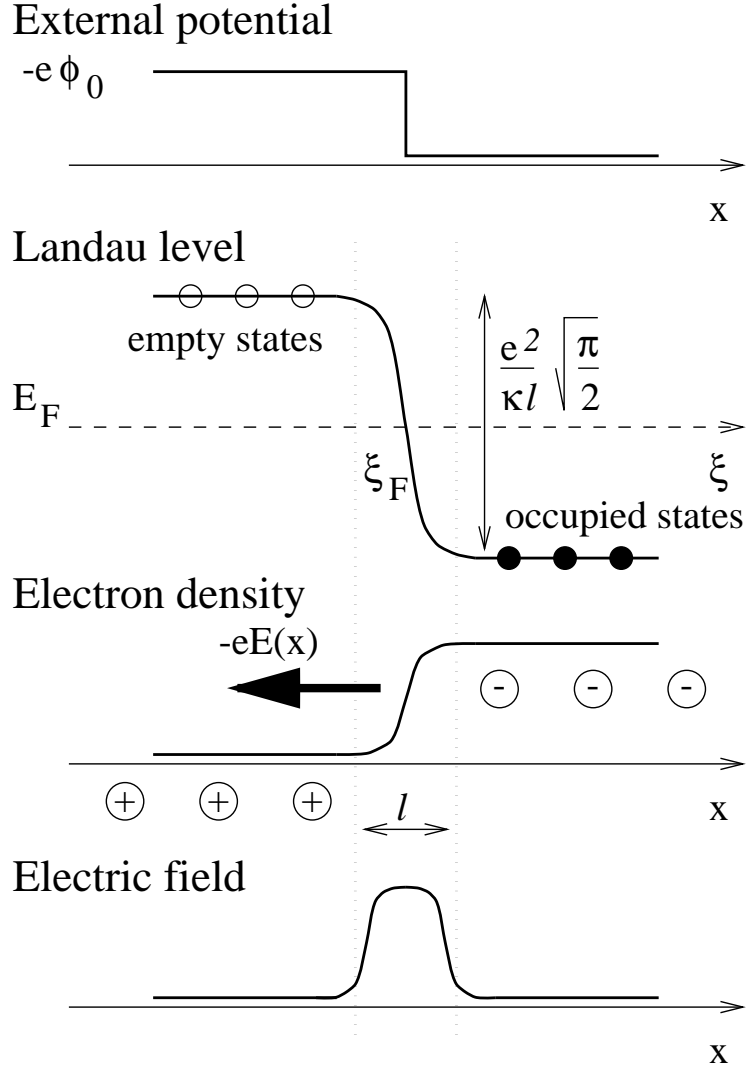}
\end{center}
\vspace{3cm}
\caption{Simplified picture of the shearing effect on the Landau levels
due to the exchange interaction.}
\end{figure}
\newpage     
\begin{figure}
\epsfxsize 10cm
\begin{center}
\epsffile{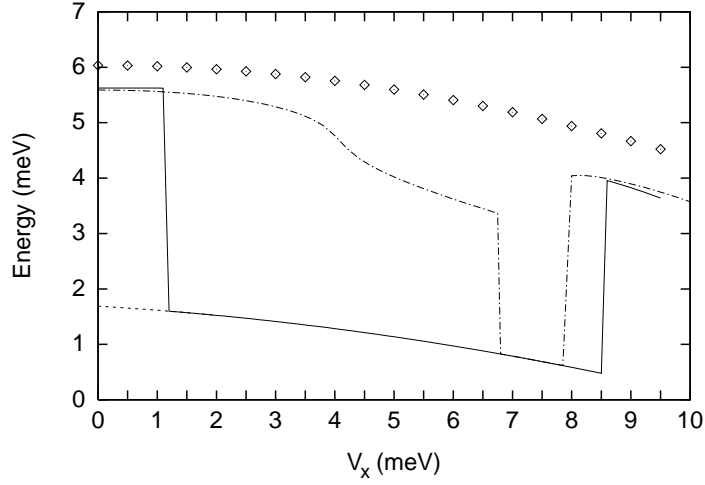}
\end{center}
\vspace{1cm}
\caption{Evolution of the energy per particle with the modulation amplitude,
in the SHFA. For $T=0.9$ K the energy goes along the solid line with 
increasing $V_x$ and returns along the dotted line.  For $T=1.3$ K the 
solid line is replaced by the dash-dotted line.  With points we show the
result of the Hartree approximation.  Other parameters are the same 
as for Fig.\ 1.}
\end{figure}
\begin{figure}
\epsfxsize 12cm
\vspace{-1cm}
\begin{center}
\epsffile{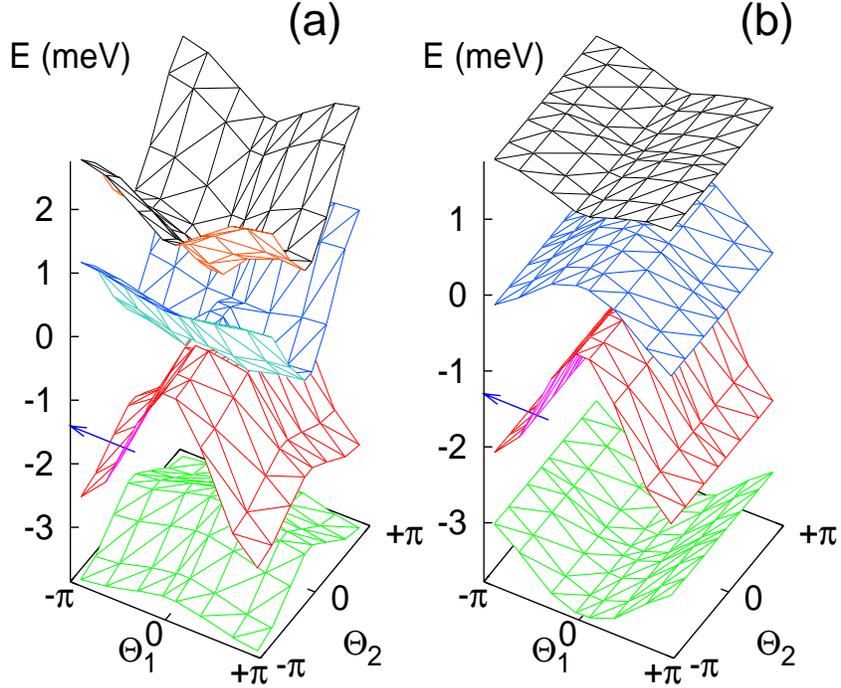}
\end{center}
\vspace{-1cm}
\caption{The self-consistent dispersion of the energy bands 
         and the Fermi level (arrow) in the first
         magnetic Brillouin zone for (a) finite external potential, and
         (b) the final result after the suppression of the potential.
         $B=1.18$ T, filling factor 1.5.}
\end{figure}
\end{document}